\documentclass[preprint,aps]{revtex4}
\usepackage{mathrsfs}

\usepackage{graphicx}
\usepackage{multirow}
\usepackage{makecell}
\usepackage{booktabs}
\usepackage{lipsum}
\usepackage{setspace}
\usepackage{bm}
\usepackage{xcolor}
\definecolor{linkwine}{RGB}{120,30,60}
\usepackage[colorlinks,linkcolor = linkwine, anchorcolor = blue,urlcolor = blue,	citecolor = blue]{hyperref}
\usepackage{float}
\usepackage{graphicx}
\usepackage{makecell}
\usepackage{lineno}

\begin{document}
%\linenumbers

\title{Anisotropic Surface State Band Splitting and Low Energy Flat Bands in 3d Correlated Topological Kondo Insulator Candidate FeSb$_2$}

\author{Ziling Cao$^{1,2,\dagger}$, Jie Pang$^{1,2,\dagger}$, Yu Xu$^{1,\dagger}$, Taimin Miao$^{1,\dagger}$, Bo Liang$^{1,2}$, Wenpei Zhu$^{1,2}$, Neng Cai$^{1,2}$, Mingkai Xu$^{1,2}$, Jumin Shi$^{1,2}$, Yingjie Shu$^{1,2}$, Yiwen Chen$^{1,2}$, Jiachen Wang$^{1,2}$, Shenjin Zhang$^{3}$, Fengfeng Zhang$^{3}$, Feng Yang$^{3}$, Zhimin Wang$^{3}$, Qinjun Peng$^{3}$, Zhihai Zhu$^{1,2}$, Xintong Li$^{1,2}$, Hanqing Mao$^{1,2}$, Guodong Liu$^{1,2}$, Zuyan Xu$^{3}$, Youguo Shi$^{1,2}$, Lin Zhao$^{1,2*}$ and X. J. Zhou$^{1,2*}$}

\affiliation{
\\$^{1}$National Laboratory for Superconductivity, Beijing National Laboratory for Condensed Matter Physics, Institute of Physics,
Chinese Academy of Sciences, Beijing 100190, China
\\$^{2}$University of Chinese Academy of Sciences, Beijing 100049, China
\\$^{3}$Technical Institute of Physics and Chemistry, Chinese Academy of Sciences, Beijing 100190, China
\\$^{\dagger}$These authors contributed equally to this work.
\\$^{*}$Corresponding author: lzhao@iphy.ac.cn and XJZhou@iphy.ac.cn
}

\date{\today}

\maketitle

\noindent {\bf{Abstract}}\\

{\bf
FeSb$_2$ is a correlated narrow-gap semiconductor that has often been discussed as a $3d$-electron Kondo insulator candidate and exhibits a low-temperature resistance plateau with possible surface-dominated conduction. We carried out a systematic high-resolution laser-based angle-resolved photoemission spectroscopy (ARPES) study of FeSb$_2$ to investigate its electronic structure. 
The surface states around the zone center show clear anisotropic splitting. 
When the temperature is lowered into the resistance plateau regime ($<6\,\mathrm{K}$), the surface states remain robust, but their photoemission peaks become much sharper and gain spectral weight. 
Two distinct flat-band-like features are observed at low energy. One is located at $\sim$127\,meV below the Fermi level, which exists only along a specific high-symmetry direction, while the other is located at $\sim$70\,meV below the Fermi level and is present along all the measured momentum cuts around the zone center. 
These results provide new information to understand the renormalization effects, the resistance plateau, and the topological nature of FeSb$_2$.
}\\


\newpage
\noindent {\bf{1. Introduction}}

\vspace{1.5mm}

Topological Kondo insulators have emerged as an important platform for studying how strong electronic correlations reshape band topology. In canonical $4f$-electron Kondo insulators, the coherent hybridization between localized $f$ moments and itinerant conduction electrons opens a narrow gap at low temperature, while spin-orbit coupling and band inversion can give rise to metallic boundary states within the bulk gap \cite{Dzero2010_TKI,FISK1996_KI,Lu2013_Correlated_TI,Dzero2016_TKI_review}. This framework has motivated extensive studies of SmB$_6$, YbB$_{12}$, Ce$_3$Bi$_4$Pt$_3$, U$_3$Bi$_4$Ni$_3$ and related narrow-gap correlated insulators, where low-temperature resistance saturation is often interpreted as a signature of residual surface conduction short-circuiting an insulating bulk \cite{Wolgast2013_SurfaceConduction_SmB6,Kim2013_SurfaceConduction_SmB6,Sato2021_SurfaceConduction_YbB12,Wakeham2016_RT_Ce3Bi4Pt3,Christopher2025_ScienceAdvances_U3Bi4Ni3}. Nevertheless, even in the prototypical $f$-electron systems, the topological origin of the observed surface states remains under active debate, as surface polarity, reconstruction, disorder, and correlation-driven in-gap states can complicate a simple topological interpretation \cite{Xu2014_TKI_SS_SmB6,Hlawenka2018_trivial_SS_SmB6,Ohtsubo2019_nontrivial_SS_SmB6,Hagiwara2016_nontrivial_SS_YbB12,Pickem2021_RT_Ce3Bi4Pt3}.

FeSb$_2$ provides a particularly intriguing $3d$-electron counterpart to this problem. As a narrow-gap correlated semiconductor, it exhibits several phenomenological similarities to canonical Kondo insulators, including activated transport, strong correlation effects, and a low-temperature resistance plateau \cite{Petrovic2005PRB_KI_FeSb2,Takahashi2011PRB_Magnetotransport_FeSb2,Homes2018SciRep_Optics_FeSb2,Xu2020PNAS_SurfaceStates_FeSb2}. These observations have motivated discussions of FeSb$_2$ as a $d$-electron Kondo-insulator candidate, yet its microscopic origin remains debated. Alternative scenarios based on itinerant narrow-gap band physics, defect-induced in-gap states, phonon-drag effects, and spin-state excitations have also been proposed \cite{Diakhate2011PRB_Thermodynamics_FeSb2,Battiato2015_FeSb2_Colossal,Du2021npj_defect_FeSb2,Li2024PNAS_FeSb2_XAS_SSE}. In this context, the origin of the low-temperature resistance plateau is a central issue. Recent transport measurements have indicated that the plateau is associated with surface-dominated conduction on top of an increasingly insulating bulk \cite{Eo2023APL_Corbino_FeSb2_FeSi,Eaton2024arXiv_Nonlocal_FeSb2}. Angle-resolved photoemission spectroscopy (ARPES) studies have directly revealed metallic surface states in FeSb$_2$\cite{Xu2020PNAS_SurfaceStates_FeSb2,Chikina2020PRR_ARPES_FeSb2}. However, whether these surface states are topologically nontrivial, reconstruction-induced, disorder-related, or manifestations of a more general correlated surface electronic structure remains unresolved \cite{Eaton2024arXiv_Nonlocal_FeSb2,Horn2026_Corbino_3Ddoping_FeSb2}. FeSb$_2$ therefore offers a timely platform for examining how narrow-gap correlation physics, surface metallicity, and possible topology are intertwined in a $3d$-electron system.

Here we investigate the electronic structure of FeSb$_2$ by high-resolution laser-based ARPES. We show that the surface states exhibit clear anisotropic band splitting and sharp photoemission peaks emerging in the low-temperature plateau regime. In addition to the flat band at the binding energy of $\sim$127\,meV, we observed another new flat band at $\sim$70\,meV below the Fermi level. These results provide new information to understand the renormalization effects, the resistance plateau and the topological nature of FeSb$_2$.

\vspace{6mm}

\noindent {\bf{2. Experiment}}

\vspace{1.5mm}

\begin{figure*}[tbp]
\begin{center}
\includegraphics[width=1\textwidth,angle=0]{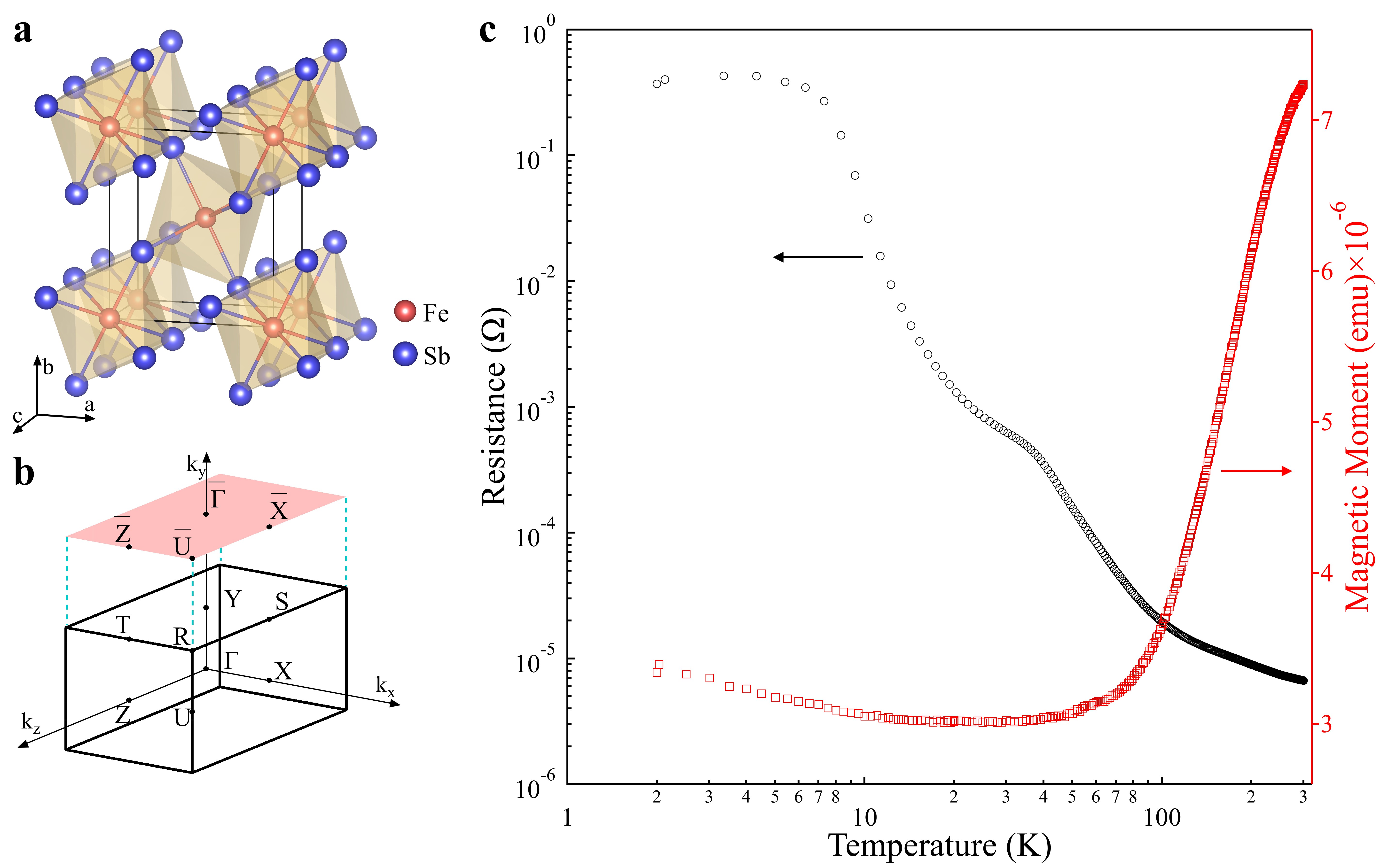}
\end{center}
\caption{{\bf Crystal Structure, Transport and Magnetic Properties of FeSb$_2$.}  
\textbf{a}, Crystal structure of FeSb$_2$,  visualized using VESTA\cite{Momma2011_VESTA3}. 
\textbf{b}, First Brillouin zone of FeSb$_2$ and the projected surface Brillouin zone of the (010) surface, with the corresponding high-symmetry points indicated. 
\textbf{c}, Temperature-dependent resistance (black curve) and magnetization (red curve) of FeSb$_2$. A pronounced resistance plateau is observed below $\sim$6\,K.
}
\label{FeSb2_intro}
\end{figure*}

High-quality single crystals of FeSb$_2$ were grown by an Sb self-flux method \cite{Petrovic2003PRB_Flux_FeSb2}. Residual Sb flux on the as-grown crystal surfaces was mechanically removed before crystallographic orientation. The crystal orientation was determined by back-reflection Laue diffraction. The oriented crystals were then cut using a diamond wire saw.  
The sample was characterized by magnetization and electrical resistance measurements (\hyperref[FeSb2_intro]{Fig.~\ref*{FeSb2_intro}c}). The magnetization was measured under an applied field of 100\,Oe with the magnetic field perpendicular to the $b$ axis. The resistance increases by approximately 5 orders of magnitude between 300\,K and 2\,K and develops a plateau below about 6\,K, consistent with previous reports for high-quality FeSb$_2$ crystals\cite{Sun2010_Dalton_FeSb2,Eo2023APL_Corbino_FeSb2_FeSi}.

\begin{figure*}[tbp]
\begin{center}
\includegraphics[width=1\textwidth,angle=0]{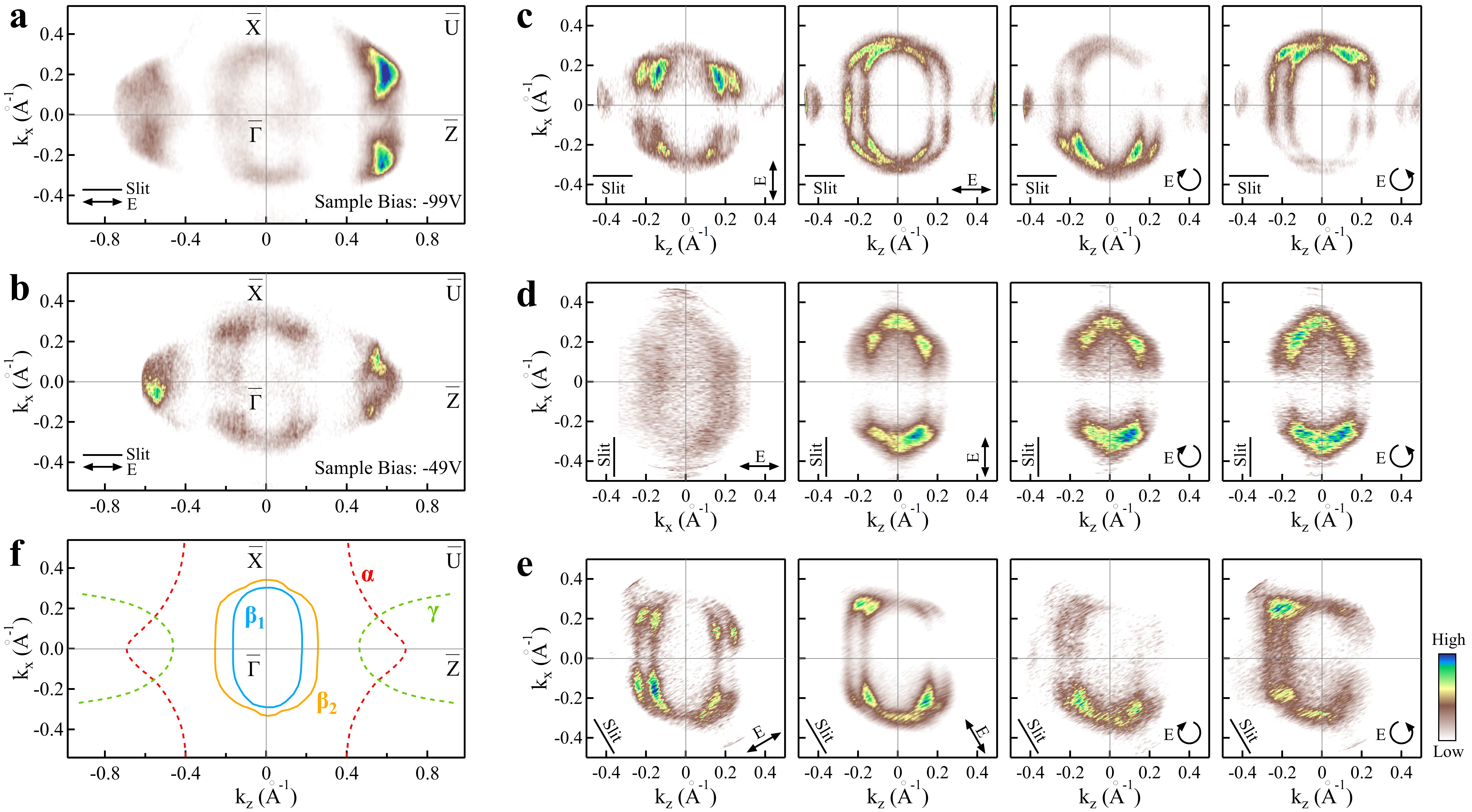}
\end{center}
\caption{{\bf Fermi surface mappings of FeSb$_2$ measured at 16\,K under different sample bias voltages\cite{Miao2026_Bias_ARPES} and different polarization geometries.} The analyzer slit orientation and the electric field, $\mathbf{E}$, of the laser light are indicated in each panel.
\textbf{a}, Fermi surface mapping of the FeSb$_2$ (010) surface measured under a sample bias voltage of -99\,V. In this case, the full 2$\pi$ solid angle of photoelectrons is collected. 
\textbf{b}, Same as \textbf{a} but measured under a sample bias voltage of -49\,V.
\textbf{c-e}, Fermi surface mappings measured under a sample bias voltage of -20\,V with two linear polarizations and two circular polarizations for the slit direction along $\overline{\Gamma}$--$\overline{Z}$ (\textbf{c}), $\overline{\Gamma}$--$\overline{X}$ (\textbf{d}) and $\overline{\Gamma}$--$\overline{U}$ (\textbf{e}), respectively.
\textbf{f}, Fermi surface of FeSb$_2$ extracted from experimental data, including two surface-state sheets $\beta_1$ and $\beta_2$ and the $\alpha$ and $\gamma$ sheets reported in Ref.~\cite{Xu2020PNAS_SurfaceStates_FeSb2}.
}
\label{FeSb2_FS}
\end{figure*}

\begin{figure*}[tbp]
\begin{center}
\includegraphics[width=1\textwidth,angle=0]{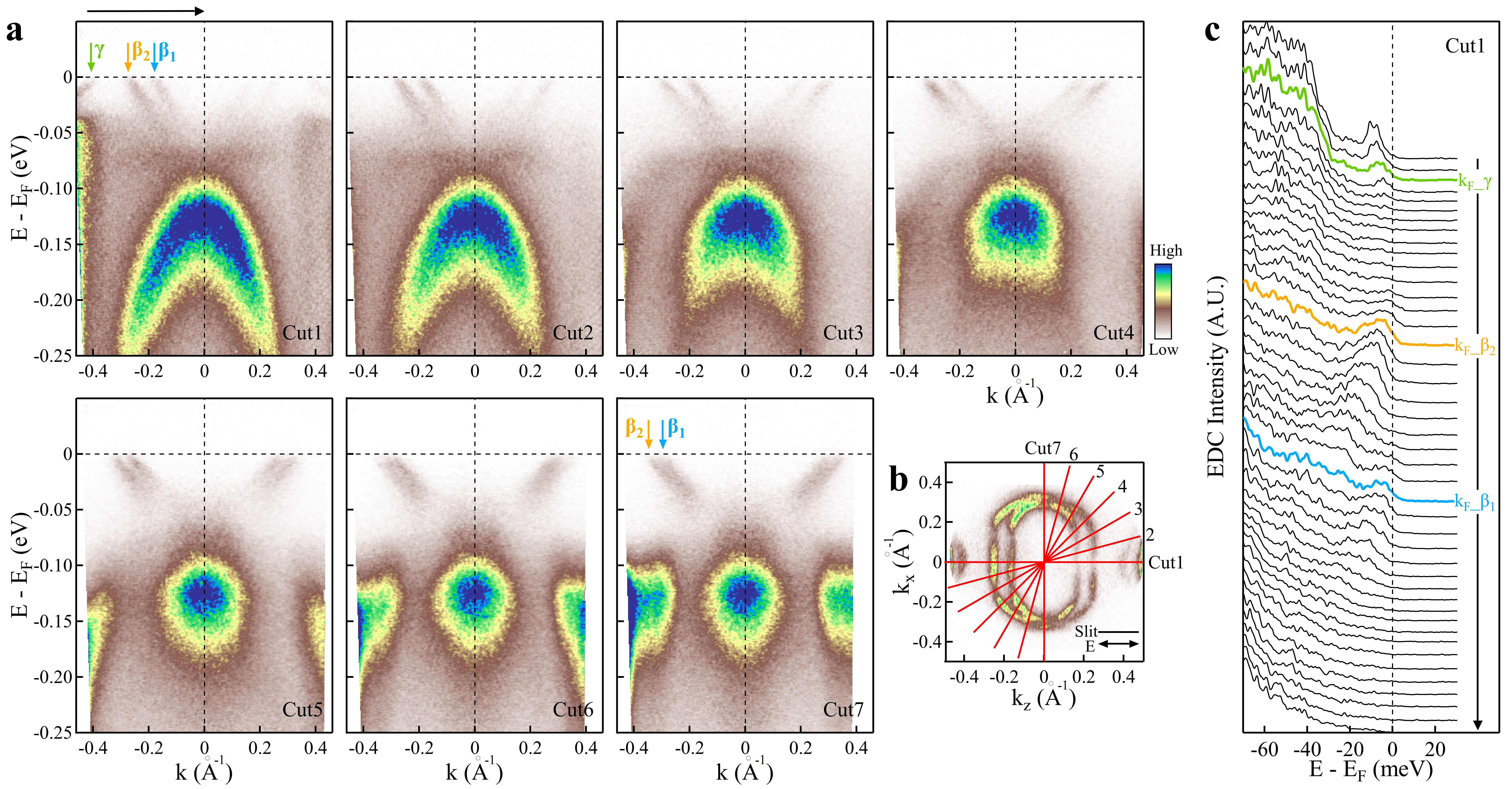}
\end{center}
\caption{{\bf Detailed Band Structures of the $\beta_1$ and $\beta_2$ bands around $\overline{\Gamma}$ measured at 16\,K under the sample bias voltage of -20\,V with the polarization geometry the same as in the second panel of Fig.~\ref{FeSb2_FS}\textbf{c}.}
\textbf{a}, Band structures measured along a series of momentum cuts centered around $\overline{\Gamma}$. The corresponding cut locations (cuts 1--7) are marked by red lines in \textbf{b}.
\textbf{b}, Fermi surface mapping of FeSb$_2$ with the momentum cuts marked.
\textbf{c}, Stacked EDCs obtained from the first panel of \textbf{a} measured along the momentum Cut1 which is along the $\overline{\Gamma}$--$\overline{Z}$ direction.
   }
\label{FeSb2_beta_Cuts}
\end{figure*}

\begin{figure*}[tbp]
\begin{center}
\includegraphics[width=1\textwidth,angle=0]{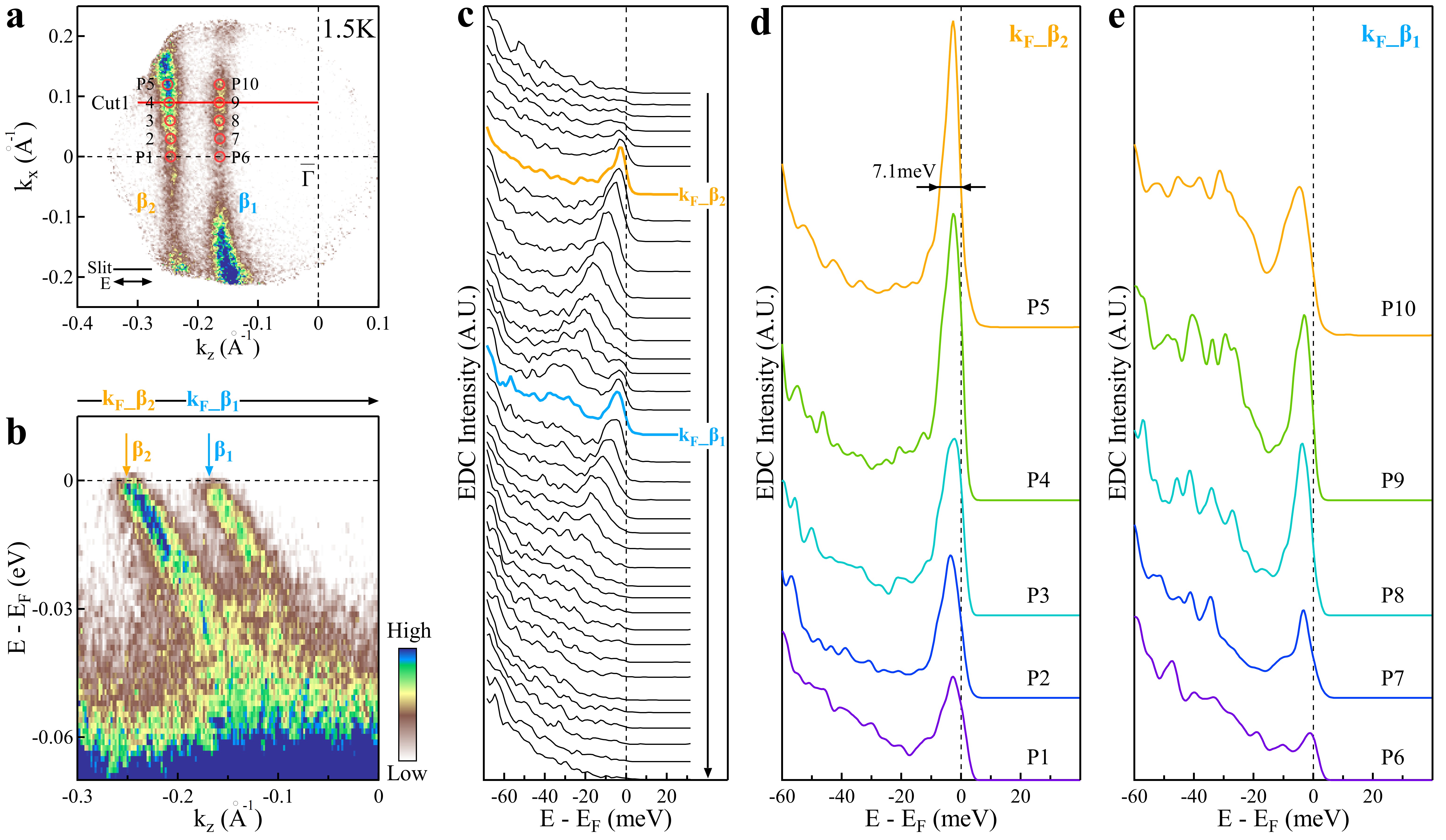}
\end{center}
\caption{{\bf Electronic structure of the $\beta_1$ and $\beta_2$ bands measured at 1.5\,K.}
\textbf{a}, Fermi surface mapping of FeSb$_2$ focusing on the $\beta_1$ and $\beta_2$ sheets.
\textbf{b}, Band structure along the momentum cut 1 marked by the red line in \textbf{a}.
\textbf{c}, Stacked EDCs obtained from \textbf{b}.
% At $k_F$, the surface-related peak is sharp and intense, with a full width at half maximum of 7.8\,meV.
\textbf{d}, EDCs along the $\beta_2$ Fermi surface. The corresponding $k_F$ points (P1--P5) are marked by red circles in \textbf{a}. For clarity, the EDCs are offset along the vertical axis.
\textbf{e}, Same as \textbf{d} but for the $\beta_1$ Fermi surface, with the $k_F$ points (P6--P10) marked by red circles in \textbf{a}.
   }
\label{FeSb2_beta_ULT}
\end{figure*}

\begin{figure*}[tbp]
\begin{center}
\includegraphics[width=1\textwidth,angle=0]{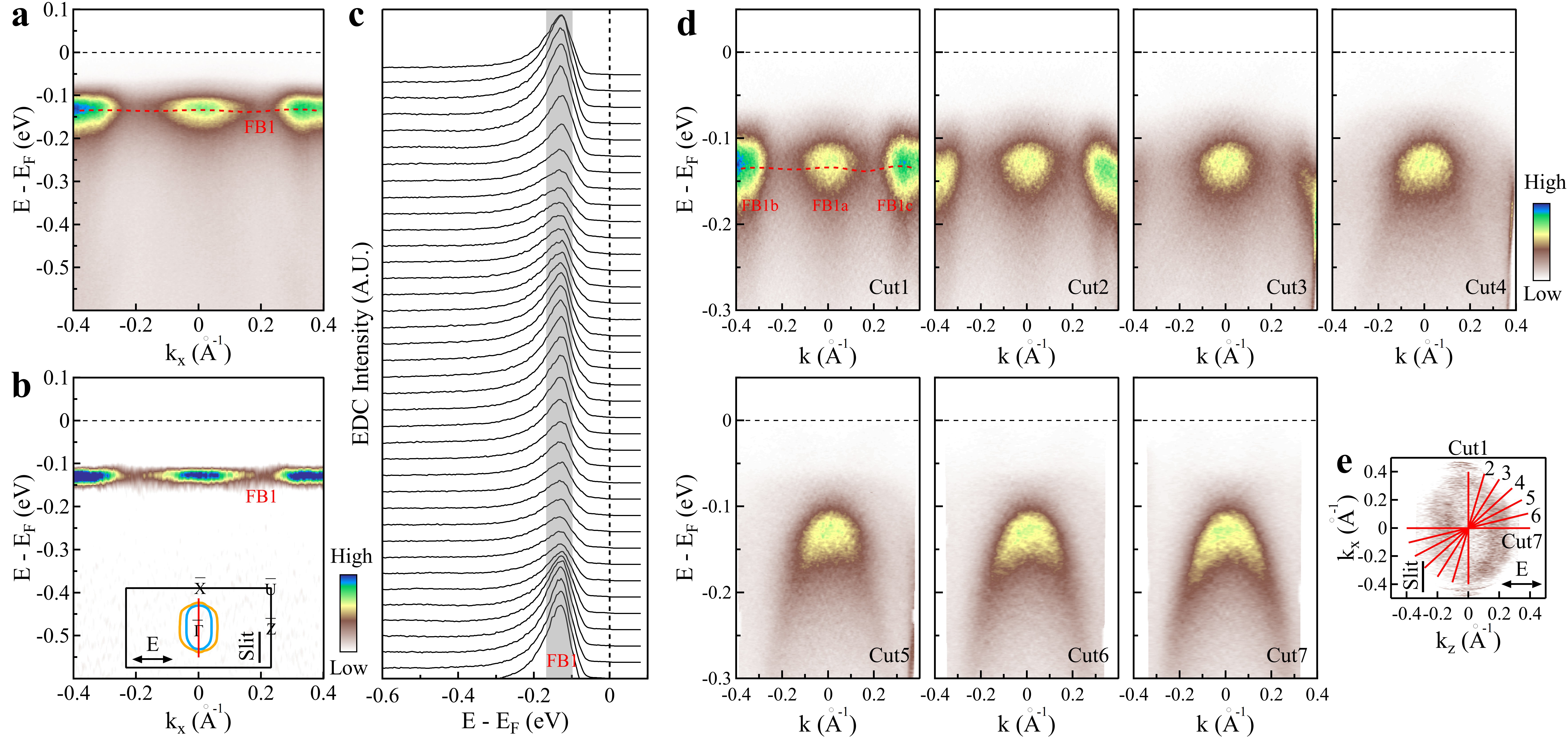}
\end{center}

\caption{{\bf Observation of a flat band FB1 at a binding energy of $\sim$127\,meV and its momentum dependence.} 
\textbf{a}, Band structure measured along the $\overline{\Gamma}$--$\overline{X}$ direction, showing a flat-band feature (FB1) at about 127 meV below $E_F$.
\textbf{b}, EDC second derivative of \textbf{a}. The measurement geometry and the location of the momentum cut (red line) are shown in the inset.
\textbf{c}, Stacked EDCs obtained from \textbf{a} where the flat band FB1 features are highlighted. 
\textbf{d}, Band structures measured along a series of momentum cuts centered around $\overline{\Gamma}$. The corresponding cut locations (cuts 1--7) are marked by red lines in \textbf{e}.
\textbf{e}, Fermi surface mapping of FeSb$_2$ with the momentum cuts marked.
}
\label{FeSb2_FB1}
\end{figure*}

\begin{figure*}[tbp]
\begin{center}
\includegraphics[width=1\textwidth,angle=0]{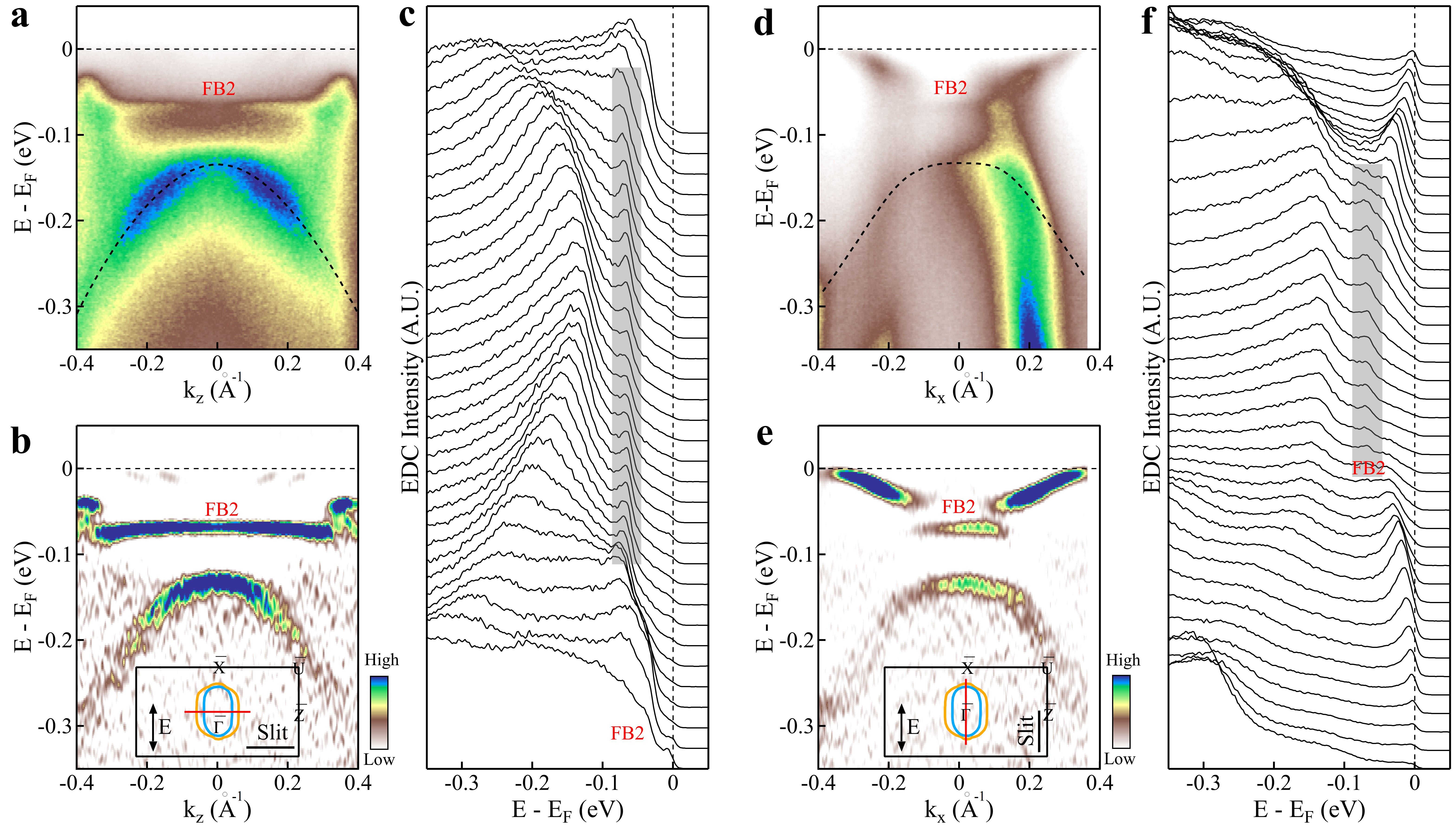}
\end{center}
\caption{{\bf Observation of a flat band FB2 at a binding energy of $\sim$70\,meV.} 
\textbf{a}, Band structure measured along the $\overline{\Gamma}$--$\overline{Z}$ direction, showing a flat-band feature (FB2) at about 70 meV below $E_F$. The bulk band is indicated by the black dashed line.
\textbf{b}, EDC second derivative of \textbf{a}. The measurement geometry and the location of the momentum cut (red line) are shown in the inset.
\textbf{c}, Stacked EDCs obtained from \textbf{a} where the flat band FB2 features are highlighted.
\textbf{d--f}, Same as \textbf{a--c} but measured along the $\overline{\Gamma}$--$\overline{X}$ direction and under the polarization geometry as marked in the inset of \textbf{e}.
}
\label{FeSb2_FB2}
\end{figure*}

ARPES measurements were carried out using laser-based systems with a photon energy of $h\nu$ = 6.994\,eV, equipped with either a hemispherical DA30L analyzer (Scienta-Omicron) (\hyperref[FeSb2_FS]{Figs.~\ref*{FeSb2_FS}, ~\ref*{FeSb2_beta_Cuts}, ~\ref*{FeSb2_FB1} and ~\ref*{FeSb2_FB2}}) or an angle-resolved time-of-flight (ARToF) analyzer (Scienta-Omicron) (\hyperref[FeSb2_beta_ULT]{Fig.~\ref*{FeSb2_beta_ULT}})\cite{Liu2008_RSI_LaserARPES,Zhou2018_RPP_LaserARPES}. 
The energy resolution was set to $\sim$1\,meV, and the angular resolution is $\sim$0.2$^\circ$ with zero sample bias. To expand the accessible momentum range, a sample-bias technique was employed during the measurements\cite{Miao2026_Bias_ARPES}.
The ARPES measurements were performed under different polarization geometries by combining various light polarizations, analyzer-slit orientations and the sample orientations.
The light polarization can vary between linear polarizations and circular polarizations. All samples were cleaved \textit{in situ} at low temperature and measured under ultrahigh vacuum with a base pressure better than $5\times10^{-11}$\,mbar. The cleavage temperature was 1.5\,K for the ARToF measurements and 16\,K for the DA30L measurements. The Fermi level ($E_F$) was 
referenced by measuring a clean polycrystalline gold which 
was electrically connected to the sample.

\vspace{4.5mm}

\noindent {\bf{3. Results and discussion}}

\vspace{1.5mm}

%Figure1: crystal structure, Brillouin zone, and transport
FeSb$_2$ crystallizes in the orthorhombic marcasite structure (space group $Pnnm$). As shown in \hyperref[FeSb2_intro]{Fig.~\ref*{FeSb2_intro}a}, each Fe atom is surrounded by six nearest-neighbor Sb atoms, forming a distorted FeSb$_6$ octahedron, while neighboring octahedra share edges along the short c axis\cite{Petrovic2003PRB_Flux_FeSb2,Petrovic2005PRB_KI_FeSb2}. The lattice constants are $a \approx 5.82$\,$~\text{\AA}$, $b \approx 6.52$\,$~\text{\AA}$ and $c \approx 3.19$\,$~\text{\AA}$. The high-symmetry points of the corresponding bulk Brillouin zone and their projections onto the (010) surface Brillouin zone are indicated in \hyperref[FeSb2_intro]{Fig.~\ref*{FeSb2_intro}b}.

% Figure 2: zone-center pocket, matrix-element dependence, and anisotropic splitting
\hyperref[FeSb2_FS]{Figure~\ref*{FeSb2_FS}} shows Fermi surface mappings of the cleaved FeSb$_2$(010) surface measured at 16\,K under different measurement conditions. 
We first measured the work function of the cleaved surface by applying a sample bias; the work function was determined to be $\Phi = 4.781$\,eV.
Then the Fermi surface mappings were measured using different sample biases and different polarization geometries.  
The application of the sample bias allows us to cover a wider momentum range. 
When the sample bias is set to -99\,V (\hyperref[FeSb2_FS]{Fig.~\ref*{FeSb2_FS}a}), the full 2$\pi$ solid angle of photoelectrons can be collected, which gives the largest possible momentum coverage by the 6.994\,eV laser excitation. 
Since the application of the sample bias deteriorates the momentum resolution, we also measured the Fermi surface with a lower sample bias of -49\,V (\hyperref[FeSb2_FS]{Fig.~\ref*{FeSb2_FS}b}) and -20\,V (\hyperref[FeSb2_FS]{Fig.~\ref*{FeSb2_FS}c-e}). 
To make use of the photoemission matrix element effects, and in particular to reveal the orbital character of the observed Fermi surface, we also measured the Fermi surface mappings under different polarization geometries combining various light polarizations, analyzer-slit orientations and the sample orientations.

From our Fermi surface measurements in \hyperref[FeSb2_FS]{Fig.~\ref*{FeSb2_FS}}, together with the band structure analysis in \hyperref[FeSb2_beta_Cuts]{Fig.~\ref*{FeSb2_beta_Cuts}} and the previous ARPES studies\cite{Xu2020PNAS_SurfaceStates_FeSb2}, we arrive at the measured Fermi surface topology of FeSb$_2$(010) shown in \hyperref[FeSb2_FS]{Fig.~\ref*{FeSb2_FS}f}. 
It consists of two small electron-like pockets around the zone center $\overline{\Gamma}$ ($\beta_1$ and $\beta_2$), an open Fermi surface along the $\overline{\Gamma}$--$\overline{X}$ direction ($\alpha$) and a hole-like pocket around $\overline{Z}$ ($\gamma$). As the bulk FeSb$_2$ is a narrow-gap semiconductor, the observed Fermi surface features are all associated with surface-related states.
When a large sample bias of -99\,V is applied (\hyperref[FeSb2_FS]{Fig.~\ref*{FeSb2_FS}a}), in the accessible momentum space, the entire $\beta$ pockets and part of the $\alpha$ and $\gamma$ features are covered. 
But the observed $\beta$ pockets are weak and broad in this case. When the sample bias is reduced to -49\,V (\hyperref[FeSb2_FS]{Fig.~\ref*{FeSb2_FS}b}), the $\beta$ pockets become slightly clearer but remain weak. Only when the sample bias is further reduced to -20\,V (\hyperref[FeSb2_FS]{Fig.~\ref*{FeSb2_FS}c-e}), the $\beta$ pockets can still be fully covered and become much sharper and stronger under proper polarization geometries, like the second panel in \hyperref[FeSb2_FS]{Fig.~\ref*{FeSb2_FS}c}. 

The $\beta$ pockets around $\overline{\Gamma}$ show a clear splitting into two branches, which we label $\beta_1$ and $\beta_2$ in \hyperref[FeSb2_FS]{Fig.~\ref*{FeSb2_FS}f}. The observed splitting is anisotropic in momentum space. The splitting is strongest along the $\overline{\Gamma}$--$\overline{Z}$ direction, but it is weakest along the $\overline{\Gamma}$--$\overline{X}$ direction. The splitting variation can also be clearly seen in the corresponding momentum dependent band structures shown in \hyperref[FeSb2_beta_Cuts]{Fig.~\ref*{FeSb2_beta_Cuts}a}. We find that, while the splitting is smallest along $\overline{\Gamma}$--$\overline{X}$, there is still a clear splitting between $\beta_1$ and $\beta_2$ along this direction, which is most clearly revealed in the band structure measured along the $\overline{\Gamma}$--$\overline{X}$ direction (last panel in \hyperref[FeSb2_beta_Cuts]{Fig.~\ref*{FeSb2_beta_Cuts}a}).
Our high-resolution laser ARPES measurements thus provide a slightly different picture of the $\beta$ pockets compared with the previous ARPES studies where the $\beta$ band splitting was not clearly resolved along the $\overline{\Gamma}$--$\overline{X}$ direction\cite{Xu2020PNAS_SurfaceStates_FeSb2}. 

% Figure 3: optimal geometry and weak 16 K EDC signature
\hyperref[FeSb2_beta_Cuts]{Figure~\ref*{FeSb2_beta_Cuts}a} shows the band structures along a series of momentum cuts centered around $\overline{\Gamma}$, measured under the optimal geometry for resolving the $\beta$ pockets. The corresponding momentum locations of the cuts are indicated in \hyperref[FeSb2_beta_Cuts]{Fig.~\ref*{FeSb2_beta_Cuts}b}. 
The splitting of the $\beta$ band into $\beta_1$ and $\beta_2$ is clearly resolved along all the cuts, with the strongest splitting along Cut1 parallel to $\overline{\Gamma}$--$\overline{Z}$ and the weakest splitting along Cut7 parallel to $\overline{\Gamma}$--$\overline{X}$.
The Fermi velocity of both $\beta_1$ and $\beta_2$ bands is similar, which is $\sim$0.53\,eV$\cdot$\AA\ along $\overline{\Gamma}$--$\overline{Z}$ (first panel in \hyperref[FeSb2_beta_Cuts]{Fig.~\ref*{FeSb2_beta_Cuts}a}) and $\sim$0.28\,eV$\cdot$\AA\ along $\overline{\Gamma}$--$\overline{X}$ (last panel in \hyperref[FeSb2_beta_Cuts]{Fig.~\ref*{FeSb2_beta_Cuts}a}). 
Considering the Fermi momentum of the two bands, this gives an effective mass of about 2.5\,$m_e$ for $\beta_1$ and 3.8\,$m_e$ for $\beta_2$ along $\overline{\Gamma}$--$\overline{Z}$, and about 8.0\,$m_e$ for $\beta_1$ and 9.1\,$m_e$ for $\beta_2$ along $\overline{\Gamma}$--$\overline{X}$. 
\hyperref[FeSb2_beta_Cuts]{Figure~\ref*{FeSb2_beta_Cuts}c} shows the stacked energy distribution curves (EDCs) for Cut1. At 16\,K, the $\gamma$, $\beta_1$ and $\beta_2$ bands all develop well-defined peaks near $E_F$ in spite of their relatively weak spectral weight. The width of the $\beta_1$ and $\beta_2$ peaks is about 10\,meV.

% Figure 4: ultralow-temperature data
To further examine the electronic structure of FeSb$_2$ in the low-temperature plateau regime, we performed laser ARPES measurements at 1.5 K. The measured Fermi surface mapping is shown in \hyperref[FeSb2_beta_ULT]{Fig.~\ref*{FeSb2_beta_ULT}a}, while \hyperref[FeSb2_beta_ULT]{Fig.~\ref*{FeSb2_beta_ULT}b} shows the band structure measured along a momentum cut parallel to $\overline{\Gamma}$--$\overline{Z}$ (Cut1). The associated stacked EDCs are presented in Fig.~\ref{FeSb2_beta_ULT}c.
Overall, the electronic structure at 1.5 K is qualitatively similar to that measured at 16 K. The two surface bands, $\beta_1$ and $\beta_2$, remain clearly visible around $\overline{\Gamma}$ and preserve the same split structure observed at higher temperature. Their spectral weight, however, is substantially enhanced at 1.5 K. In the stacked EDCs for Cut1 (\hyperref[FeSb2_beta_ULT]{Fig.~\ref*{FeSb2_beta_ULT}c}), the corresponding peaks become markedly sharper and stronger near $E_F$.
To examine this behavior more systematically, we extracted a series of EDCs at the $k_F$ points on the $\beta_2$ (\hyperref[FeSb2_beta_ULT]{Fig.~\ref*{FeSb2_beta_ULT}d}) and $\beta_1$ (\hyperref[FeSb2_beta_ULT]{Fig.~\ref*{FeSb2_beta_ULT}e}) pockets. Although the peak intensities vary from point to point because of the matrix element effects, the peaks are consistently much narrower at 1.5\,K, with a width of about 7.1\,meV. 
Despite the pronounced low-temperature sharpening and spectral-weight enhancement, no clear gap opening is resolved near the peak positions. Within the momentum region around $\overline{\Gamma}$ probed here, cooling into the low-temperature plateau regime is therefore not accompanied by an obvious band reconstruction or any similarly abrupt change in the electronic structure. Instead, the observed evolution is more consistent with enhanced coherence of pre-existing states rather than a low-temperature electronic phase transition\cite{Pickem2021_RT_Ce3Bi4Pt3,Eaton2024arXiv_Nonlocal_FeSb2}.

% Figure 5: FB1 and its relation to the previously reported bulk alpha band
Now we turn to the low-energy flat bands observed in FeSb$_2$. 
\hyperref[FeSb2_FB1]{Fig.~\ref*{FeSb2_FB1}a} shows a clear observation of a weakly dispersive feature, labeled FB1, at $\sim$127\,meV below $E_F$ in the band structure measured along the $\overline{\Gamma}$--$\overline{X}$ direction. This FB1 band exhibits a pronounced matrix-element dependence and is observed most clearly under the measurement geometry indicated in the inset of \hyperref[FeSb2_FB1]{Fig.~\ref*{FeSb2_FB1}b}.
The feature becomes more pronounced in the EDC second derivative image shown in \hyperref[FeSb2_FB1]{Fig.~\ref*{FeSb2_FB1}b}. 
\hyperref[FeSb2_FB1]{Fig.~\ref*{FeSb2_FB1}c} shows the stacked EDCs for the same cut, where the FB1 feature is resolved as a clear peak at $\sim$127\,meV below $E_F$. Within the measured momentum range, the energy position of FB1 varies by no more than 6\,meV, indicating a weakly dispersive character. We note that the EDC peaks are broad, which is $\sim$100\,meV in width. The energy position of this FB1 feature and its weakly dispersive character closely resemble those of the strongly renormalized band reported in previous ARPES and dynamical mean-field theory (DMFT) studies\cite{Xu2020PNAS_SurfaceStates_FeSb2,Chikina2020PRR_ARPES_FeSb2}. 

To study the momentum dependence of FB1, we measured band structures along a series of momentum cuts centered around $\overline{\Gamma}$, as shown in \hyperref[FeSb2_FB1]{Fig.~\ref*{FeSb2_FB1}d}. As shown in the second derivative image in \hyperref[FeSb2_FB1]{Fig.~\ref*{FeSb2_FB1}a} and the first panel in \hyperref[FeSb2_FB1]{Fig.~\ref*{FeSb2_FB1}d}, the FB1 feature appears to be composed of three sub-features, one central feature FB1a and two side features FB1b and FB1c. 
When the momentum cut is slightly rotated away from the $\overline{\Gamma}$--$\overline{X}$ direction (second panel in \hyperref[FeSb2_FB1]{Fig.~\ref*{FeSb2_FB1}d} for Cut2), the FB1a feature remains at essentially the same energy, while the FB1b and FB1c features show a clear dispersion. When the momentum cuts are further rotated (third to seventh panels in \hyperref[FeSb2_FB1]{Fig.~\ref*{FeSb2_FB1}d} for Cut3 to Cut7), the FB1b and FB1c features disappear in the covered momentum region. 
These results indicate that the overall flat-band-like FB1 feature exists over a narrow momentum region around $\overline{\Gamma}$--$\overline{X}$ direction.

% Figure 6: FB2 and the bulk beta band
\hyperref[FeSb2_FB2]{Figure~\ref*{FeSb2_FB2}} shows the first observation of another flat-band-like feature, labeled FB2, at $\sim$70\,meV below $E_F$. 
\hyperref[FeSb2_FB2]{Fig.~\ref*{FeSb2_FB2}a} presents the band structure measured along the $\overline{\Gamma}$--$\overline{Z}$ direction. A flat-band-like feature is clearly observed at $\sim$70\,meV below $E_F$, labeled as FB2.
This FB2 feature is more clearly revealed in the EDC second derivative image shown in \hyperref[FeSb2_FB2]{Fig.~\ref*{FeSb2_FB2}b}.
\hyperref[FeSb2_FB2]{Fig.~\ref*{FeSb2_FB2}c} shows the stacked EDCs for the same cut, where the FB2 feature is resolved as a clear peak at $\sim$70\,meV below $E_F$. The FB2 feature exists within a finite momentum range between [-0.33, 0.33]\,\AA$^{-1}$ along $\overline{\Gamma}$--$\overline{Z}$. 
Within the momentum range, the energy position of FB2 varies by no more than 5\,meV, indicating an extremely weak dispersion. Different from the FB1 feature, the EDC peaks of FB2 are sharp, with a width of about 15 meV. 

\hyperref[FeSb2_FB2]{Figure~\ref*{FeSb2_FB2}d-f} show the band structure, EDC second derivative image and stacked EDCs for a cut along the $\overline{\Gamma}$--$\overline{X}$ direction. The FB2 flat band feature is also clearly observed along this direction at the same energy position of $\sim$70\,meV below $E_F$. The momentum range of FB2 along $\overline{\Gamma}$--$\overline{X}$ is about [-0.13, 0.13]\,\AA$^{-1}$.
Different from the FB1 feature, the FB2 flat band feature is present along all the momentum cuts around the $\overline{\Gamma}$ point.

% Discussion
An important question raised by the present results is about the origin of the observed surface states and their relation to the low-temperature resistance plateau. 
So far, there have been no band structure calculations that can fully capture the observed surface states\cite{Homes2018SciRep_Optics_FeSb2,Chikina2020PRR_ARPES_FeSb2}. 
The topological nature of the observed surface states also needs further investigation. 
A related question concerns how the resistance plateau is reflected in the low-energy states observed by ARPES. As shown in \hyperref[FeSb2_beta_ULT]{Fig.~\ref*{FeSb2_beta_ULT}}, within the momentum window around $\overline{\Gamma}$, cooling into the plateau regime does not induce an abrupt band reconstruction, a clear gap opening, or a distinct new in-gap feature. Instead, the dominant changes are a sharpening of the pre-existing $\beta$ branches and an enhancement of their spectral weight. This behavior shows that the plateau regime is associated with the low-temperature strengthening of pre-existing states, rather than with a sudden electronic reconstruction. 

\vspace{6mm}

\noindent {\bf{4. Summary}}

\vspace{1.5mm}

High-resolution laser-based ARPES measurements were carried out on FeSb$_2$. 
The surface states around $\overline{\Gamma}$ exhibit clear anisotropic band splitting. 
When the sample is cooled down into the low-temperature plateau regime, the surface states remain robust, but their photoemission peaks become much sharper and gain spectral weight. 
We observed two distinct flat-band-like features at low energy. One is located at $\sim$127\,meV below the Fermi level, which exists only along $\overline{\Gamma}$--$\overline{X}$, while the other is located at $\sim$70\,meV below the Fermi level and is present along all the measured momentum cuts around $\overline{\Gamma}$.
The present results support a multicomponent picture of FeSb$_2$, with surface $\beta$ branches, flat low-energy features, and correlation-driven renormalization contributing to the plateau-regime electronic structure. 
Because the present ARPES data mainly probe the vicinity of $\overline{\Gamma}$, more subtle low-temperature changes elsewhere in the surface Brillouin zone, such as near $\overline{Z}$ or $\overline{U}$, cannot be excluded. ARPES spectral weight alone also cannot determine the contribution of the surface-related channel to the total electrical conduction. 
Further progress will require wider momentum-space mapping, spin-resolved ARPES, and more theoretical efforts.

\vspace{6mm}

\noindent {\bf Acknowledgements}

\vspace{1.5mm}

\noindent This work is supported by the National Key Research and Development Program of China (Grant No. 2024YFA1408301, 2021YFA1401800, 2022YFA1604200, 2022YFA1403900, 2023YFA1406002, 2023YFA1406103 and 2024YFA1408100), the National Natural Science Foundation of China (Grant No. 12488201, 12374066, 12374154 and 12494593), CAS Superconducting Research Project (Grant No. SCZX-0101), Quantum Science and Technology National Science and Technology Major Project of China (Grant No. 2021ZD0301800), and Synergetic Extreme Condition User Facility (SECUF).

\vspace{6mm}

\noindent {\bf Author Contributions}

\vspace{1.5mm}

\noindent X.J.Z., L.Z. and Z.L.C. proposed and designed the research. J.P., Y.G.S. and Z.L.C. contributed to crystal growth, characterization and sample preparation. Z.L.C. and Y.X. performed the ARPES measurements. Z.L.C., Y.X., T.M.M., B.L., W.P.Z., N.C., M.K.X., J.M.S., Y.J.S., Y.W.C., J.C.W., S.J.Z., F.F.Z., F.Y., Z.M.W., Q.J.P., Z.H.Z., X.T.L., H.Q.M., G.D.L., Z.Y.X., L.Z. and X.J.Z. contributed to the development and maintenance of the laser ARPES systems. X.J.Z. and Z.L.C. analyzed the data and wrote the manuscript. All authors participated in the discussion and commented on the manuscript.

\vspace{3mm}

\newpage
{\bf{References}}\\

\newpage

\end{document}